\documentclass[%
 reprint,
 amsmath,amssymb,
 aps,
nofootinbib]{revtex4-2}

\usepackage{graphicx}
\usepackage{caption}
\usepackage{subcaption}
\usepackage{ragged2e} 
\DeclareCaptionJustification{justified}{\justifying}
\usepackage{dcolumn}
\usepackage{bm}
\usepackage{float}
\usepackage{xcolor, soul}
\usepackage{bm}
\usepackage{physics}
\usepackage{epstopdf}
\usepackage{amsmath}
\usepackage[english]{babel}
\usepackage[autostyle, english = american]{csquotes}

\MakeOuterQuote{"}

\begin{document}

\title{Calibration of electric fields in low-frequency off-resonant Rydberg receivers}%

\author{Baran Kayim}
\email{baran.kayim@gtri.gatech.edu}
\author{Michael A. Viray}
\author{David S. La Mantia}
\author{Daniel Richardson}
\thanks{Current address: University of Colorado Boulder, Boulder, Colorado 80301, USA}
\author{James Dee}
\author{Ryan S. Westafer} 
\author{Brian C. Sawyer}
\author{Robert Wyllie}
\email{robert.wyllie@gtri.gatech.edu}
\affiliation{Georgia Tech Research Institute, Atlanta, Georgia 30332, USA}

\date{\today}
\begin{abstract} We present results on Rydberg atom-based electric field sensing in the range of 1~kHz~-~300~MHz, using a three-photon Rydberg excitation scheme and a transverse electromagnetic (TEM) line waveguide to apply low-frequency rf fields to the cell. Measurements of low-frequency screening in quartz and sapphire vapor cells show excellent agreement with a phenomenological model of the effective vapor cell material properties based on an electrical 2-port measurement of the TEM line. We achieve a best noise-equivalent field of 106(4)~$\mathrm{\frac{\mu V}{m \sqrt{Hz}}}$ at 300 MHz and characterize noise-equivalent fields in the ultra-low to very-low frequency (ULF-VLF) band.
\end{abstract}

\maketitle
\section{\label{sec:intro}Introduction}
Rydberg atoms~\cite{gallagherbook} are a promising platform for SI-traceable, self-calibrated, and broadband passive radio-frequency (rf) receivers, with high sensitivities owing to large transition dipole moments and dynamical polarizabilities in the Rydberg manifold~\cite{gallagherbook,holloway2017atom,fancollision}. Rydberg atomic receivers have also demonstrated capabilities such as phase-sensitive~\cite{simons_mixerac_2019,jingheterodyne} and  polarization-sensitive~\cite{sedlacek2013polarization,bao2016polarization} detection, in addition to angle-of-arrival rf measurements with both sensor arrays and individual vapor cells~\cite{robinson2021aoa,richardson2025study,schlossberger2025angle}. Typical detection experiments include spectroscopic probing using Autler-Townes (AT) splitting~\cite{fancollision} or off-resonant AC Stark shifts~\cite{anderson2017continuous,li2022rydberg}, and the electric field sensitivities of these schemes can be further enhanced with homodyne~\cite{kumar2017atom} and heterodyne~\cite{jingheterodyne,gordon2019weak} schemes using an additional local oscillator (LO) field. 

\begin{figure}[]
    \centering
    \includegraphics[width = 0.98\linewidth]{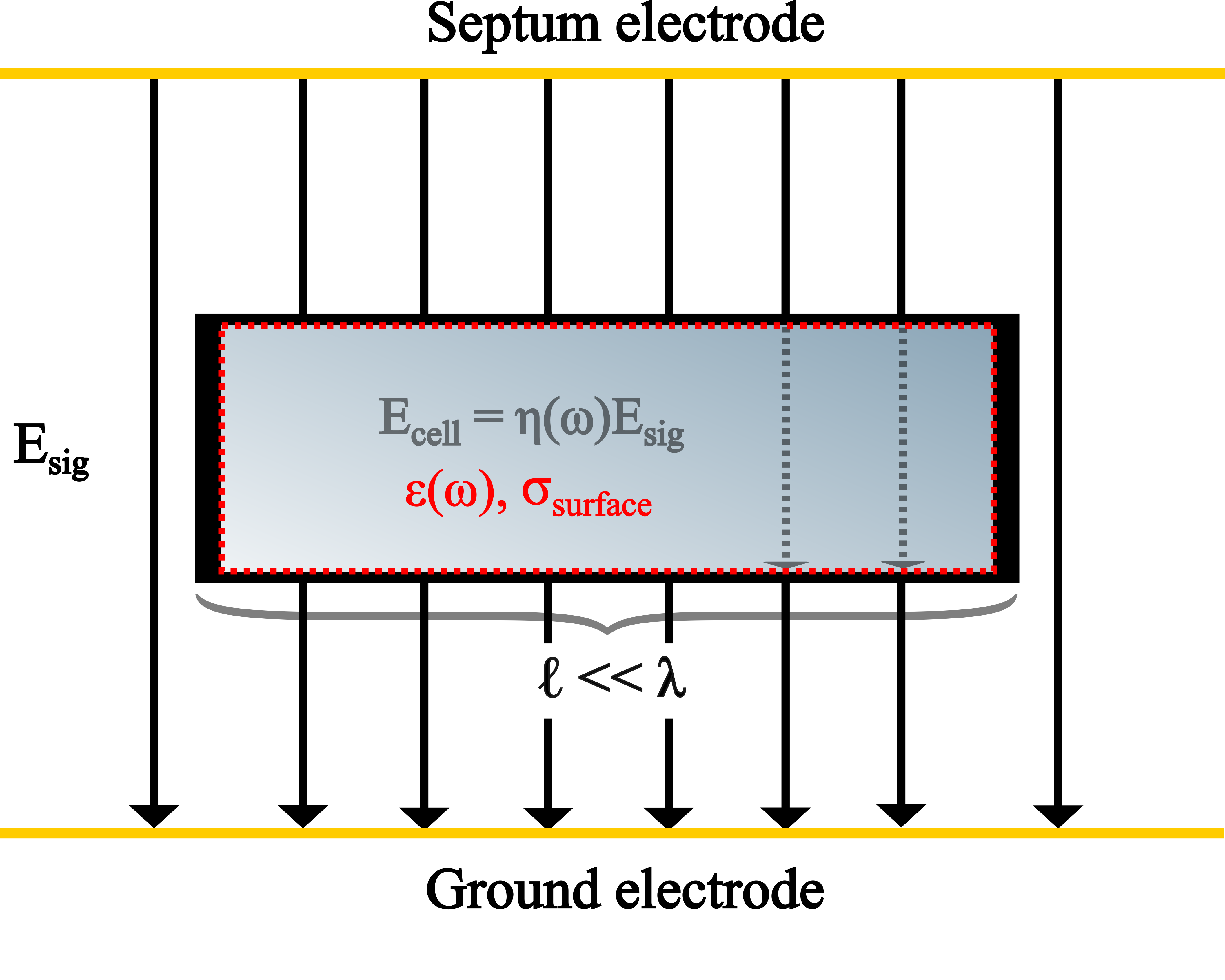}
    \caption{Depiction of the vapor cell shielding effect, where the frequency-dependent bulk dielectric constant $\epsilon(\omega)$ and the sheet conductivity of the inner surface $\sigma_{surface}$ induced by adsorption of the alkali atoms conductively shields electric field $E_{sig}$ applied by a pair of external electrodes to yield a reduced field amplitude $E_{cell}~=~\eta(\omega)E_{sig}$ for $\eta(\omega)~\leq~1$. In the frequency range covered in this paper, vapor cell length $\ell \ll \lambda$ for signal field wavelength $\lambda$}
    \label{fig:0}
\end{figure}

Resonant transitions between Rydberg states typically allow for detection ranging from hundreds of GHz to tens of MHz~\cite{brown2023very,vsibalic2017arc}. However, resonant detection of frequencies below the VHF band ($<$300~MHz) has proven challenging, requiring either higher principal quantum number states or high angular momentum states. Furthermore, both resonant and off-resonant Rydberg receivers from the ultra-low frequency to very-high frequency (ULF to VHF) range (300~Hz - 300~MHz) experience conductive screening from atoms adsorbed on the inner surfaces of vapor cells (as illustrated in Fig.~\ref{fig:0}). The magnitude and frequency dependence of this conductive screening effect varies with vapor cell material, wall temperature, and vapor pressure~\cite{bouchiat1999electrical,jau2020vapor,danielTEMpaper,ma2025study}. The increase in the effective sheet conductivity of the vapor cell's inner surface from these adsorbates has a high-pass filtering effect, causing a frequency-dependent attenuation of the electric field reaching the atomic vapor~\cite{jau2020vapor,adams2019rydberg,PhysRevA.75.062903}. At low frequencies where the vapor cell dimensions ($\ell$) are deeply sub-wavelength $(\ell \ll \lambda)$, we can model this behavior with a frequency-dependent shielding factor $\eta(\omega)$ (as seen in Fig.~\ref{fig:0}):
\begin{equation}
\label{eq:1}
E_{cell} = \eta(\omega)E_{sig}
\end{equation}

\noindent where $E_{cell}$ is the electric field amplitude internal to the vapor cell, $E_{sig}$ is the external signal electric field incident on the vapor cell, and $\eta(\omega) \leq 1$. This is distinct from the effect in vapor cells for $\ell \gg \lambda$, wherein spatial variation of the rf electric field amplitude is caused by partial standing waves formed inside the vapor cell~\cite{fan2015effect}. 

In this paper we present a calibrated Rydberg atom measurement of the screening effect that is in excellent agreement with a recently-developed material extraction model for dispersive effects in vapor cells~\cite{danielTEMpaper}. This model is based on a completely independent electromagnetic measurement of the vapor cell's two-port parameters. It is important for the benchmarking of Rydberg receivers as fieldable electric field sensors to both measure and model the factor $\eta(\omega)$, since knowledge of this screening factor is critical to accurately report low-frequency noise-equivalent field in reference to the incoming field amplitude, as opposed to a self-calibrated atomic measurement ignoring the screening effect.

To characterize the screening effect, we implement a three-photon Rydberg receiver scheme in rubidium vapor~\cite{brown2023very}, utilizing a transverse electromagnetic (TEM) waveguide to apply uniform 1~kHz~-~300~MHz rf electric fields to the room-temperature vapor cell at the Rydberg atoms~\cite{danielTEMpaper}. For continuously-tunable field detection in the VHF band, we measure the Stark shift of an electromagnetically induced
transparency (EIT)~\cite{boyd2008nonlinear,boller1991observation} transmission peak in response to an off-resonant signal field applied to the waveguide. We achieve a best noise-equivalent field of 106(4)~$\mathrm{\frac{\mu V}{m \sqrt{Hz}}}$ at 300 MHz using the 55F$_{7/2}$ state. Throughout this paper, reported noise-equivalent fields include any relevant vapor cell screening effects.

This paper is structured as follows: in Sec.~\ref{sec:exp} we present our experimental apparatus and a characterization of our TEM waveguide. In Sec.~\ref{sec:results} we present both our measured low-frequency screening coefficients and our noise-equivalent field results, and briefly summarize the Debye material model and methods to extract the Debye material properties that we have presented more fully elsewhere~\cite{danielTEMpaper}. Finally, in Sec.~\ref{sec:out}, we compare our noise-equivalent field results to those achieved in the literature for similar frequency ranges and conclude with an outlook for future work.

\section{\label{sec:exp}Experiment}
\begin{figure}[t]
    \centering
    \begin{subfigure}[b]{\linewidth}
        \centering
        \includegraphics[width = \linewidth]{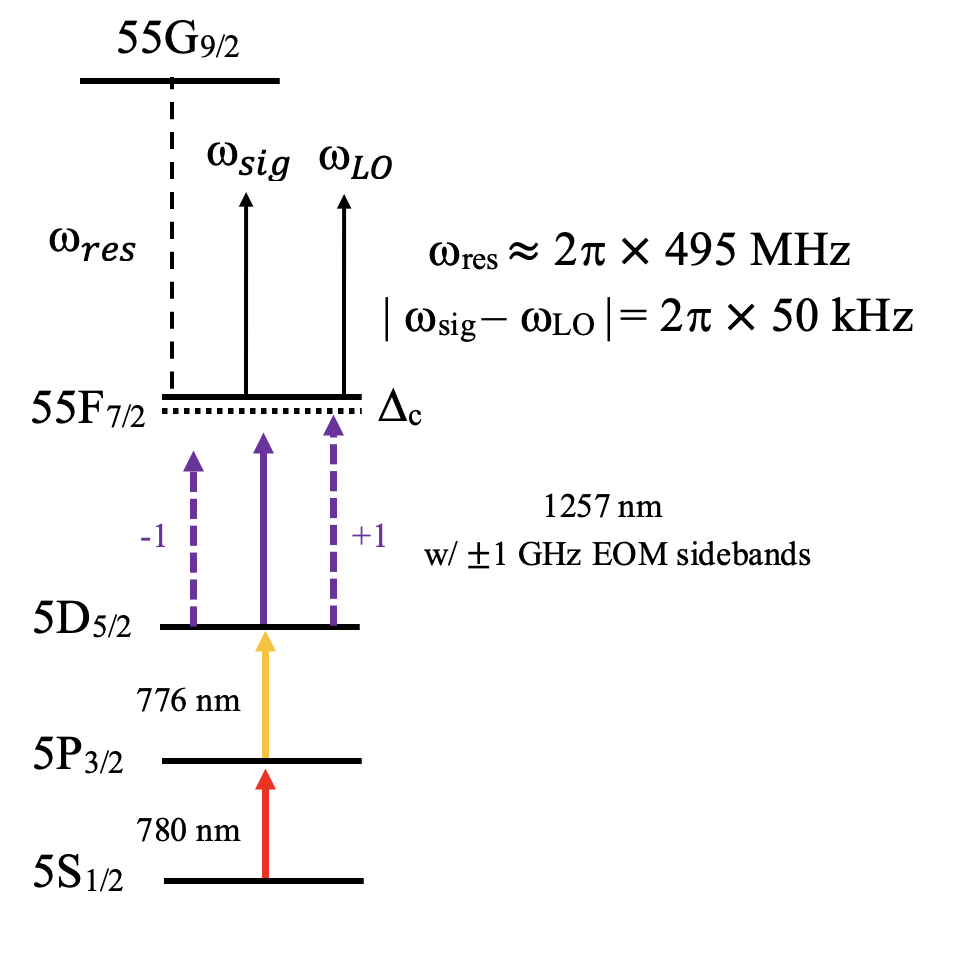}
    \end{subfigure}

    \begin{subfigure}[b]{\linewidth}
        \centering
        \includegraphics[width = \linewidth]{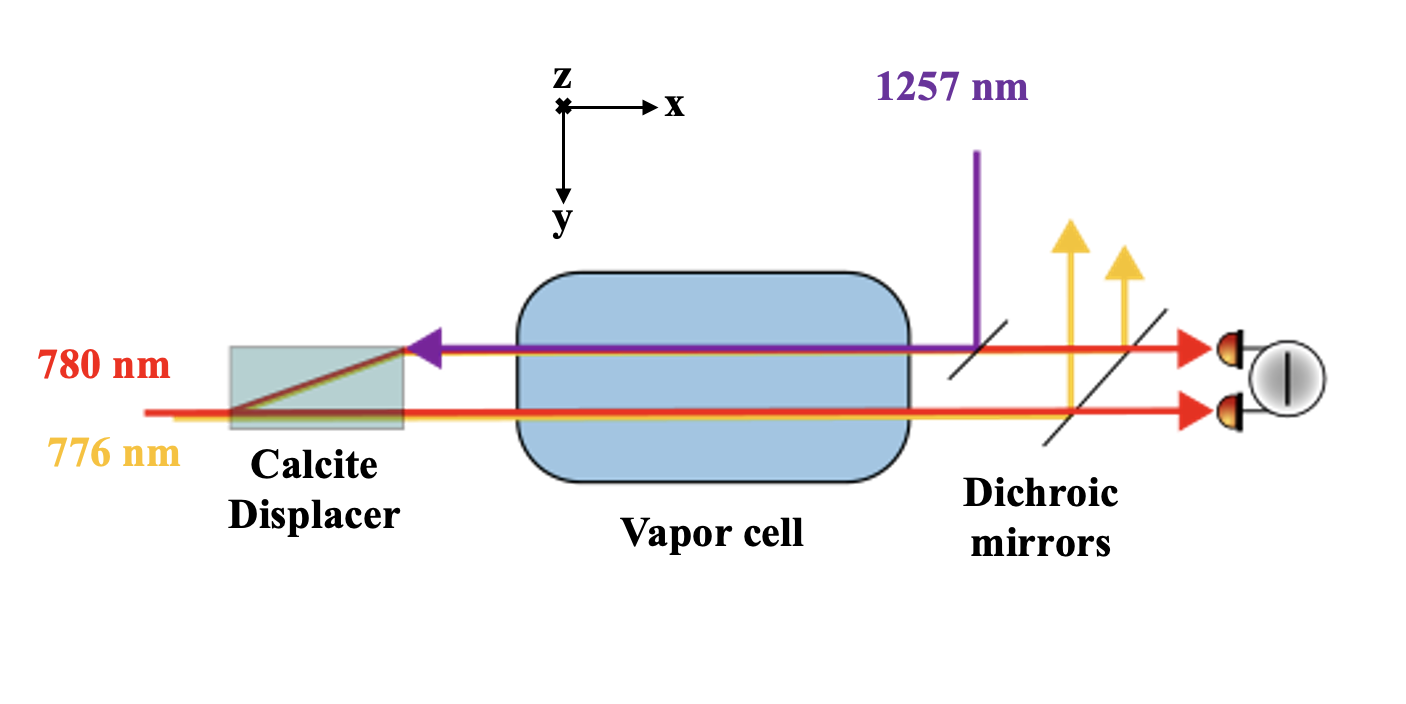}
    \end{subfigure}
    
    \caption{(Top) Level diagram for optical excitation in the off-resonant detection scheme, where three optical transitions connect the Rb ground state to the Rydberg state $55F_{7/2}$. An off-resonance signal field with frequency $\omega_{sig}$ and an off-resonance local oscillator with frequency $\omega_{LO}$ incident on the Rydberg atoms induce a heterodyne beatnote $\abs{\omega_{sig} - \omega_{LO}}$ modulating the transmitted probe intensity. The nearest resonant microwave transition is at approximately $\omega_{res} \approx 2\pi \times 495$ MHz. Also displayed are EOM sidebands induced on the 1257~nm coupler laser for scanning across resonance. (Bottom) Diagram of laser beam paths through vapor cell, showing region where co-propagating 780 and 776~nm lasers overlap with the 1257~nm coupler.}
    \label{fig:1}
\end{figure}

We present results from two vapor cells: (1) a 6.9~cm long, 16~mm internal diameter cylindrical quartz vapor cell\footnote{Thorlabs part number: GC19075-RB87} with wall thickness 1.5~mm filled with isotopically pure $^{87}$Rb, and (2) a 1.4~cm side-length cubic sapphire vapor cell\footnote{Manufactured by Japan Cell Inc.} with a natural-abundance mixture of $^{85}$Rb (72.2$\%$) and $^{87}$Rb (27.8$\%$). In the sapphire cell, we address transitions in $^{85}$Rb. For both cells, we utilize a three-photon excitation scheme (Fig.~\ref{fig:1}) to access the 55F$_{7/2}$ state in Rb, similar to previous work~\cite{brown2023very}. The laser wavelengths 780 nm, 776 nm, and 1257 nm, respectively couple state transitions $\ket{5S_{1/2}, F = 2}$ $\rightarrow$ $\ket{5P_{3/2}, F' = 3}$ $\rightarrow$ $\ket{5D_{5/2}, F'' = 4}$ $\rightarrow$ $\ket{55F_{7/2}}$ for $^{87}$Rb and $\ket{5S_{1/2}, F = 3}$ $\rightarrow$ $\ket{5P_{3/2}, F' = 4}$ $\rightarrow$ $\ket{5D_{5/2}, F'' = 5}$ $\rightarrow$ $\ket{55F_{7/2}}$ for $^{85}$Rb. The detuning of the 1257~nm coupler laser is represented by $\Delta_c$, while the other two lasers are tuned onto resonance. The principal quantum number n~=~55 is chosen empirically as a Rydberg state balancing high atomic polarizability against minimal EIT lineshape broadening due to factors such as collisional broadening or stray DC electric field gradients within the vapor cell. We define a figure of merit given by the ratio of EIT peak height to full width at half-maximum ratio for several states of differing principal quantum number. 

\begin{figure}[t]
    \centering
    \begin{subfigure}[b]{\linewidth}
        \centering
        \includegraphics[width = \linewidth]{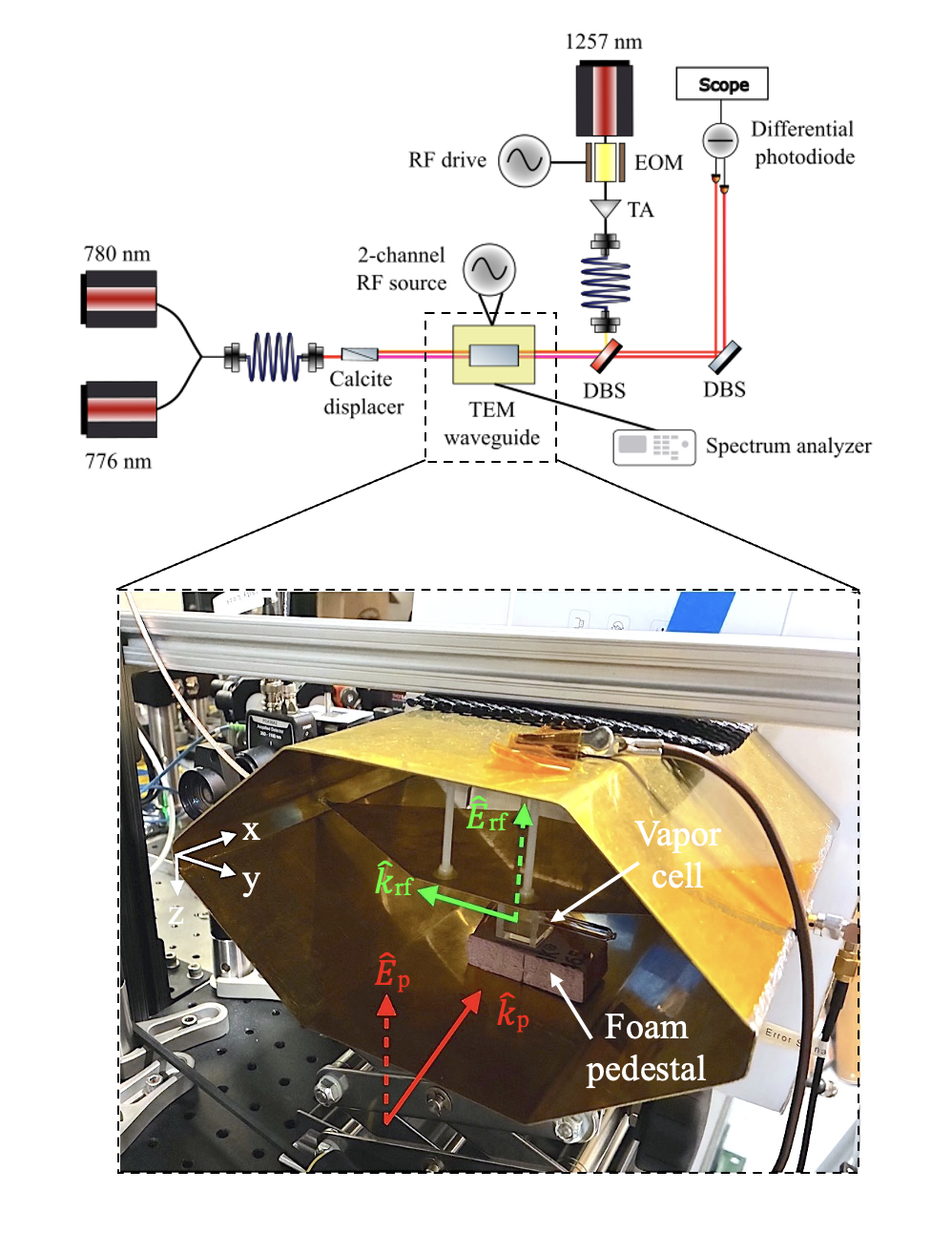}
    \end{subfigure}
    
    \caption{Simplified experimental schematic. PD: photodiode; DBS: dichroic beam splitter; AOM: acousto-optical modulator; EOM: electro-optical modulator. In the middle is an image of the TEM waveguide with a sapphire vapor cell mounted under the central conductor, with wavevectors of probe and rf fields and the coordinate system used in HFSS simulation}
    \label{fig:2}
\end{figure}

\begin{figure}[t]
    \centering
    \includegraphics[width = 0.95\linewidth]{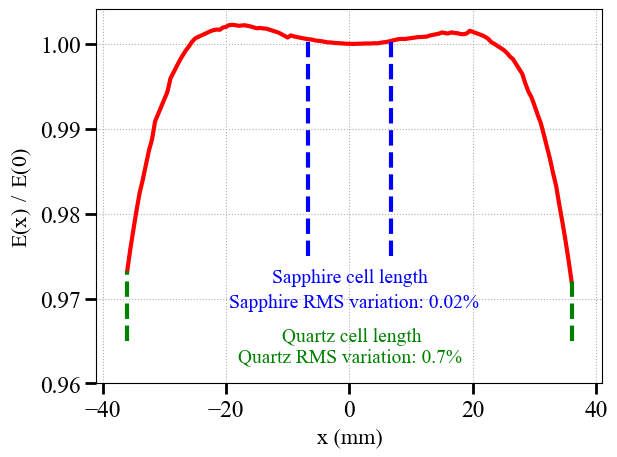}
    \caption{Simulated $E(x)/E(0)$ for $E(x)$ averaged across the cross-sectional beam overlap area against x-position for both vapor cells, derived from HFSS simulations of empty waveguide for a signal field at 300 MHz. Also included (inset text) are the matching RMS variances in electric field for both quartz and sapphire vapor cells.}
    \label{fig:3}
\end{figure}

The optical layout is shown in Fig.~\ref{fig:2}. The vapor cells are optically addressed through an excitation column by the co-propagating 780~nm and 776~nm laser beams and a counter-propagating 1257~nm laser beam, which we found resulted in the best figure-of-merit on-resonance EIT lineshapes. The 780~nm, 776~nm, and 1257~nm $\frac{1}{e^2}$ laser beam diameters are measured to be 1.09~mm,  1.16~mm, and 1.44~mm respectively, with respective powers (Rabi frequencies) 14.7~$\mu$W (2$\pi \times$7.55~MHz), 401.6~$\mu$W (2$\pi \times$9.15~MHz), and 453~mW (2$\pi \times$9.45~MHz). The region where the lasers overlap acts as the interaction volume in the atomic vapor. The lasers are locked to a commercial wavemeter, which is calibrated to a second 780~nm laser locked via Doppler-free polarization rotation spectroscopy to a separate Rb reference vapor cell~\cite{wieman1976doppler}. The 1257~nm laser is locked at a point $\sim$1~GHz red-detuned from the $5D_{5/2} \rightarrow 55F_{7/2}$ transition frequency, then passed through a fiber-coupled electro-optic phase modulator (EOM) driven near 1~GHz. We used the~+1~sideband for EIT spectroscopy, as shown in Fig.~\ref{fig:1}. We then swept the frequency of the rf source driving the EOM to tune the coupling laser across three-photon resonance. All experimental control, including laser frequency locking and data collection, was handled with the Labscript control system~\cite{starkey2013scripted}, and the electronics in the experiment were synchronized to a 10 MHz rubidium frequency reference.

We used a differential photodetector to measure the transmission difference between a 780~nm probe and reference beam. The 780~nm reference beam, as well as the co-propagating 776~nm beam, was displaced 4~mm by a calcite displacer and propagated through a volume in the vapor cell not overlapping with the 1257~nm beam. The phase-synchronous signal and local oscillator fields are applied with two ports of a single rf signal generator using a combiner prior to the TEM waveguide port. The TEM guide, shown in Fig.~\ref{fig:2}, was mounted on the optical table with the top of the vapor cell 2~cm below the center conductor. We use a custom waveguide, rather than a commercial TEM cell, to ensure an accurate rf-model could be made of the cell and to optimize electric field uniformity across the vapor cell. The custom TEM waveguide used in this experiment was designed to provide uniform fields and maintain TEM mode purity up to 300 MHz~\cite{rotunnoinhomogeneity,fan2015effect,danielTEMpaper}. Field inhomogeneity in the vapor cell can broaden the EIT line~\cite{rotunnoinhomogeneity}. Electromagnetic simulation of the empty waveguide (shown in Fig.~\ref{fig:3}) indicates a maximum RMS variation in electric field amplitude of 0.7\% across the 6.9~cm laser interaction length at 300 MHz in the quartz vapor cell, and 0.02\% across the 1.4~cm laser interaction length at 300 MHz in the sapphire vapor cell, where the RMS variation is defined as
\begin{equation*}
    \sqrt{\frac{\sum_x E^2(x) - E^2(0)}{N \times E^2(0)}}
\end{equation*}

\noindent where $N$ is the number of x-position indices simulated in the waveguide interior, $E(x)$ is the electric field at each x-position averaged across y- and z-axes in the cross-sectional area of the overlapping beams as shown in Fig.~\ref{fig:1}, and $E(0)$ is the electric field at the center of either vapor cell. It should be noted that because the variation in the quartz cell is not Gaussian, there are excursions away from uniformity near the edges of the excitation volume which fall well outside this 0.7\%. The ability to apply uniform and well-modeled fields within the waveguide permits measurements of the screening from the vapor cell by applying a single effective field amplitude across the volume of the vapor cell.

\section{\label{sec:results}Results}
\subsection{Effective Material Measurement}
In this section we will briefly summarize the pertinent results of another recent publication concerning the modeling of our TEM waveguide and electric field screening in vapor cells~\cite{danielTEMpaper}, which is relevant for independently verifying the low-frequency shielding effect measured by comparing the atomic response to the expected free-space field. In~\cite{danielTEMpaper}, causal dispersive effects due to the vapor cell were modeled—using commercial software Ansys HFSS—as a Debye relaxation with an additional conductivity term, which assumes the dielectric relaxation response can be modeled  as an ideal noninteracting group of electric dipoles~\cite{PhysRevB.78.045205,DebyeBook}. This was motivated by prior literature, where an electrical conductivity of the vapor cell walls had been identified in silica glass as adsorbed alkali metal in dynamic equilibrium with the vapor cell's inner surface for glasses of particular binding energy and wall temperature~\cite{bouchiat1999electrical}. The entire thin glass body was modeled in~\cite{danielTEMpaper} with an effective complex permittivity including sheet conductivity, which is represented by the expression:
\begin{equation}
\epsilon(\omega) = \epsilon_{\infty} + \frac{\epsilon_s - \epsilon_{\infty}}{1 + i\frac{\omega}{\omega_0}} - \frac{i\sigma}{\epsilon_0\omega}
\end{equation}

\noindent where $\epsilon_{\infty}$ represents the infinite-frequency limit of permittivity, $\epsilon_{s}$ represents the zero-frequency limit, $\omega$ is the driving frequency, $\omega_0$ is the dipole relaxation rate, and $\sigma$ is the sheet conductivity of the surface. The first two terms represent effects from the bound charges, while the conductivity term represents the free charges. An optimization procedure reliant on matching the modeled and measured 2-port parameters of the waveguide with and without the vapor cell using differing Debye model values was used to find the values that best modeled the S$_{11}$ and S$_{12}$ measurements. Examples for several vapor cell materials and alkali atoms are presented in~\cite{danielTEMpaper}. This effective material model could then be used in an HFSS model of the vapor cell in the TEM guide to simulate the effective conductive screening from the vapor cell. We note this model is independent of Rydberg measurements and provides an independent estimate of the shielding factor, $\eta(\omega)$. A central result of this paper is the excellent agreement between these two screening estimates.

\subsection{Field Calibrations}
For electric field calibrations with atomic spectra, we initially model the total electric field incident on the atomic vapor as a sum of $j$ different field components with their own polarization ($\hat{\epsilon}_j$), field amplitude ($E_j$), frequency ($\omega_j$), and phase ($\phi_j$):
\begin{equation}
\vec{E}_{tot} = \sum_{j}\hat{\epsilon}_j E_j \cos(\omega_j t + \phi_j)
\end{equation}

We can then express the second-order Stark shift of an atomic energy level $\ket{a}$ $(\Delta U_a)$ as:
\begin{equation*}
\Delta U_a(t) = \frac{1}{i \hbar}\sum_{b\neq a}H'_{ab}(t)e^{i\omega_{ab}t}\int^t_0 H'_{ba}(t')e^{i\omega_{ba}t'}dt'
\end{equation*}

\noindent where $\omega_{ab}$ represents the transition frequency between state $\ket{a}$ and state $\ket{b}$. Using the interaction Hamiltonian:
\begin{equation*}
  \begin{gathered}
    H_{ab}'(t) = -\sum_{j}\bra{a}\vec{\mu} \cdot \hat{\epsilon}_j\ket{b} E_j \cos(\omega_j t + \phi_j)
  \end{gathered}
\end{equation*}

\noindent where $\vec{\mu}$ represents the induced dipole moment from the dominant external electric field, we expand the Stark energy shift $(\Delta U_a)$ to:
\begin{equation*}
    \begin{gathered}
\Delta U_a(t) = \frac{1}{i \hbar}\sum_{b\neq a}\sum_{j,k}\bra{a}\vec{\mu} \cdot \hat{\epsilon}_j\ket{b} E_j \cos(\omega_j t + \phi_j)e^{i\omega_{ab}t} \\
\times \int^t_0\bra{b}\vec{\mu} \cdot \hat{\epsilon}^*_k\ket{a}E_k \cos(\omega_k t' + \phi_k)e^{i\omega_{ba}t'}dt'
    \end{gathered}
\end{equation*}

\noindent using an additional index $k$. We consider the combination of independent static and dynamic electric fields, such as may result from an uncontrolled inner-cell environment and an applied external field.  Then, for $j, k \in \{0, 1\}$, $\omega_0 = 0$, $\omega_1 = \omega$, and $\omega^2_{ba} \gg \omega^2$ for signal field frequency $\omega$ such that the total applied field can be modeled as the combination of an rf signal field ($\vec{E_1}$) and a static electric field component ($\vec{E_{0}}$) accounting for uncontrolled fields interior to the vapor cell, the frequency shift is:
\begin{multline*}
    \begin{gathered}
\Delta f_a = \Re{\Delta U_a(t)}/2\pi\hbar = \frac{-1}{ 2\pi\hbar^2}\left[\sum_{b\neq a}\frac{E_0^2 \abs{\bra{a}\vec{\mu} \cdot \hat{\epsilon}_0 \ket{b}}^2}{\omega_{ba}} \right.\\ \left. +~E_1^2\abs{\bra{a}\vec{\mu} \cdot \hat{\epsilon}_1 \ket{b}}^2 \frac{\cos^2(\omega t + \phi_1)}{\omega_{ba}}\right.
\\ \left.+~2E_0E_1 \Re{\bra{a}\vec{\mu} \cdot \hat{\epsilon}_0 \ket{b} \bra{b}\vec{\mu} \cdot \hat{\epsilon}_1^* \ket{a}} \frac{\cos(\omega t + \phi_1)}{\omega_{ba}}\right]
    \end{gathered}
\end{multline*}

\noindent Assuming that the quantization axis is set by the polarization of the lasers, which are parallel to the direction of the rf signal field such that $\bra{a}\vec{\mu} \cdot \hat{\epsilon}_0 \ket{b} = \bra{a}\vec{\mu} \cdot \hat{\epsilon}_1 \ket{b}\cos(\theta)$ for angle $\theta$ between fields $\vec{E_{0}}$ and $\vec{E_{1}}$, the total Stark shift is given by:
\begin{multline} \label{eq:4}
\Delta f_a = -\frac{1}{2}\alpha\left[E_0^2 \cos^2(\theta) + E_1^2 \cos^2(\omega t + \phi_1) \right.\\
+\left. 2E_0 E_1 \cos(\omega t + \phi_1) \cos(\theta) \right]
\end{multline}
\begin{equation*}
    \begin{gathered}
\alpha = \frac{1}{\pi \hbar^2}\sum_b \frac{\abs{\bra{a}\vec{\mu} \cdot \hat{\epsilon}_1 \ket{b}}^2}{\omega_{ba}}
    \end{gathered}
\end{equation*}

\noindent The quantity $\alpha$ is the scalar polarizability. In this work, the  Alkali Rydberg Calculator~\cite{vsibalic2017arc} was used to calculate $\alpha$.

Based on the above analysis we conducted atomic calibrations in two separate frequency regimes, detailed in the two following subsections. 

\subsubsection{Calibrations for $\omega \geq 2\pi\times1$~MHz}
At signal frequencies $\geq 1$~MHz, one can resolve AC Stark shifts with moderate applied electric fields in both sapphire and fused silica vapor cells. In this instance, where $\omega > 2\pi f_{BW}$ for instantaneous atomic bandwidth $f_{BW} \approx 150$~kHz, the time-dependent Stark shift oscillating faster than $f_{BW}$ is time-averaged in the atomic response from the probe transmission. Then, the expression for the AC Stark shift in Equation~(\ref{eq:4}) reduces to:
\begin{equation}
\Delta f_a = -\frac{1}{2}\alpha E_1^2.
\label{eq:5}
\end{equation}

Here, we measured the EIT peak shift as a function of $E_1$, recording changes in the 780~nm probe beam absorption. We did this by frequency modulating 1257~nm EOM rf source at a 49~kHz dither frequency around $\approx$1 GHz. We then monitored the out-of-phase output of a phase-sensitive lock-in amplifier referenced to the dither frequency to recover a dispersive signal centered on the EIT resonance. The dither frequency was chosen so that it was not an integer or half-integer multiple of any of our chosen signal frequencies, and because it was within the typical EIT response bandwidth $f_{BW}$. 

\begin{figure}[t]
    \centering
    \includegraphics[width = 0.9\linewidth]{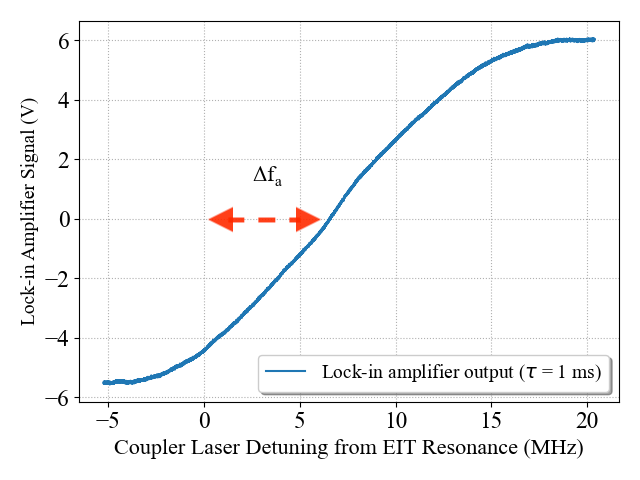} 
    \includegraphics[width = 0.9\linewidth]{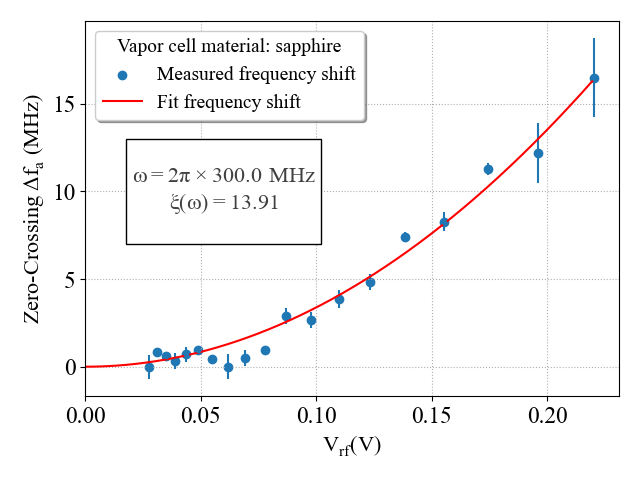}
    \caption{(Top) Plot showing lock-in amplifier (time constant $\tau$ = 1 ms) output and corresponding zero-crossing for $\omega = 2\pi \times 300$~MHz, yielding a 5.58~MHz shift for $V_{rf} = 0.12$~V ($E_1 = 1.67$~V/m) (Bottom) Frequency shift of zero-crossing versus $V_{rf}$ for $\omega = 2\pi \times 300$~MHz and overlaid fit according to Equation~(\ref{eq:5}), in addition to signal field slope parameter $\xi$.}
    \label{fig:5}
\end{figure}

Figure~\ref{fig:5} (top), shows data from a typical 1257~nm scan across the resonance. The position of the EIT resonance, at the zero crossing of the measurement, was tracked with increasing rf RMS voltage ($V_{rf}$), as measured at the output of the TEM waveguide by a spectrum analyzer. The resultant frequency shifts of the zero crossing $\Delta f_a$ for each measured $V_{rf}$ were fit according to Equation~(\ref{eq:5}), as shown in Fig.~\ref{fig:5} (bottom). This procedure was repeated for each rf frequency $\omega$. The figure shows the expected quadratic scaling of the frequency shift with $V_{rf}$, where the fit parameters of the quadratic fit allow us to calibrate applied electric field to voltage at the waveguide.

\subsubsection{Calibrations for $\mathit{\omega < 2\pi\times1}$~MHz}
\begin{figure}[]
    \centering
    \includegraphics[width = 0.9\linewidth]{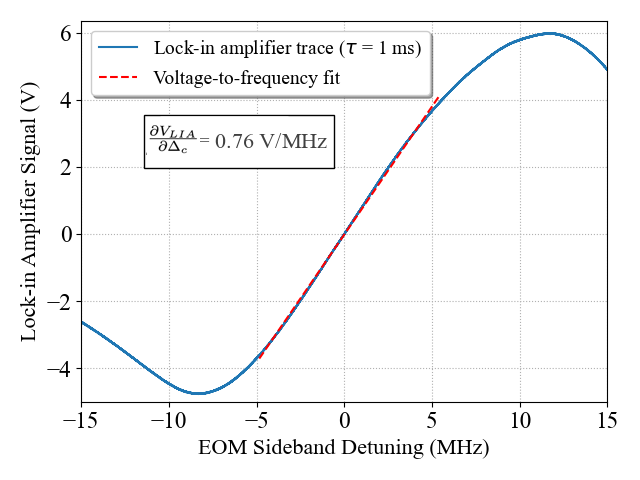} 
    \includegraphics[width = 0.9\linewidth]{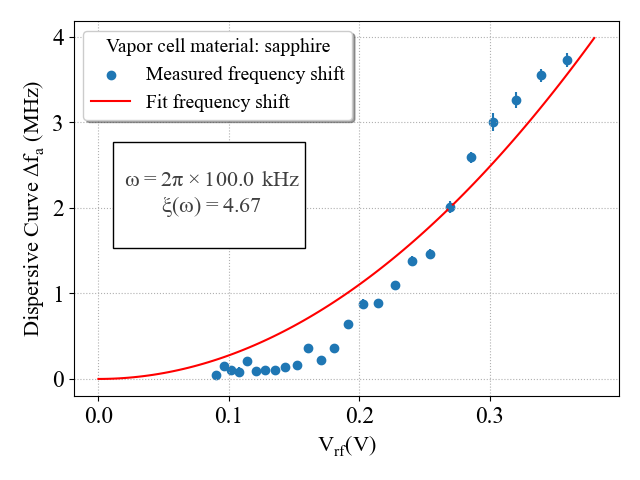}
    \includegraphics[width = 0.9\linewidth]{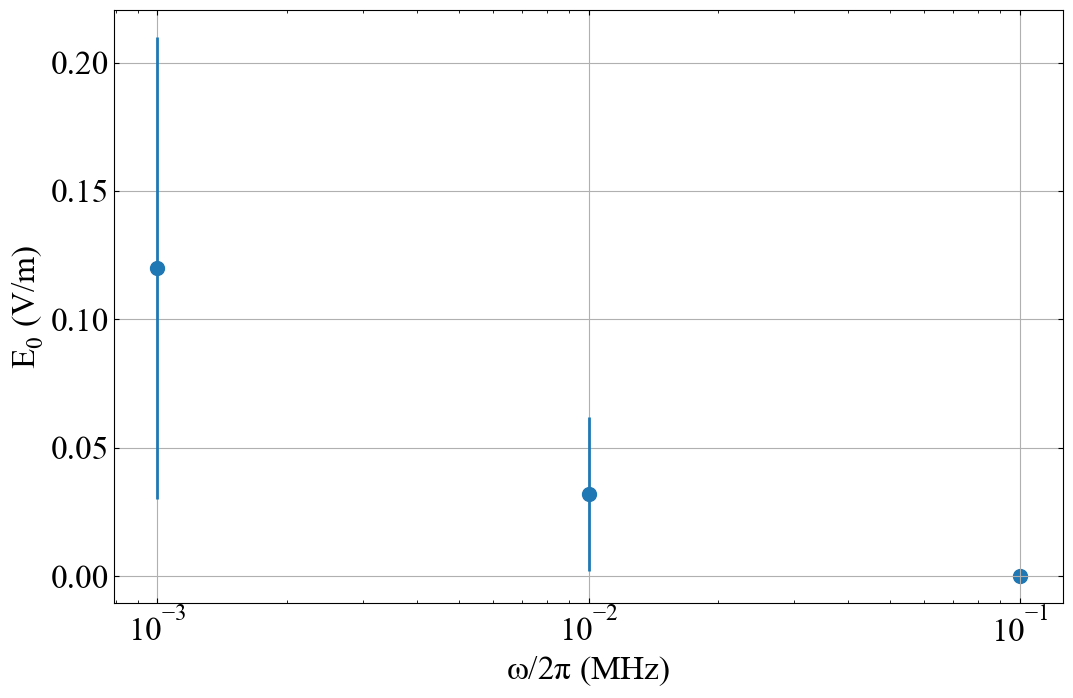}
    \caption{(Top) Plot showing lock-in amplifier (time constant $\tau$ = 1 ms) output from 780~nm laser probe transmission and graphic of how $\Delta f_a$ is calculated for $\omega = 2\pi \times 1$~MHz, with slope of $\frac{\partial V_{LIA}}{\partial \Delta_c}$. (Middle) Frequency shift of $V_{LIA}$ versus $V_{rf}$ for $\omega = 2\pi \times 100$~kHz and overlaid fit according to Equation~(\ref{eq:6}), in addition to signal field slope parameter $\xi$. (Bottom) Static bias field amplitude $E_0$ for signal frequencies 1~kHz, 10~kHz, and 100~kHz as found through calibration fits according to Equation~(\ref{eq:6}), with error bars determined by covariance of the fit parameter}
    \label{fig:6}
\end{figure}

For lower frequencies the Stark shift varies slowly and the AC Stark expression is no longer appropriate. Due to the screening effect, we found it easier to use a modified atomic calibration procedure. First, the region centered about resonance was fit to a linear curve with a given slope $\frac{\partial V_{LIA}}{\partial \Delta_c}$ (as shown in Fig.~\ref{fig:6}). Then, the 1257~nm coupler laser's EOM sideband was locked at a point blue-detuned from the EIT resonance in the linear region of the dispersive curve whilst maintaining the dither, and the rf source generating the signal field was set to the desired calibration signal frequency. The voltage value of the lock-in amplifier trace was then measured as the applied $V_{rf}$ was swept, resulting in an expression for $\frac{\partial V_{LIA}}{\partial V_{rf}}$. Then, using ${\frac{\partial V_{LIA}}{\partial \Delta_c}}^{-1}\frac{\partial V_{LIA}}{\partial V_{rf}}~=~\frac{\partial \Delta_c}{\partial V_{rf}}$, a calibration could be derived for the Stark shift in terms of $V_{rf}$. 

In this lower-frequency regime, $\omega < {2\pi f_{BW}}$ for instantaneous atomic bandwidth $f_{BW} \approx 150$~kHz and the energy shift of the dressed states is no longer time-averaged in the atomic response. Instead, the Stark shift of the probe transmission closely follows the instantaneous value of the incoming fields. Therefore, it is important for this calibration method to ensure that the demodulation and filtering from the lock-in amplifier only preserves time-varying terms proportional to $E_1$ such that Equation~(\ref{eq:4}) can still be used for fitting atomic calibrations. For the input signal to the lock-in amplifier, the probe transmission at a given coupler detuning $\Delta_c$ is related to the first-order energy shift of the dressed Rydberg state~\cite{jingheterodyne}. For small deviations from the initial setpoint of the coupler laser where the modulation of the EOM sideband and the Stark shift are presumed not to sample the nonlinearity of the EIT profile, the instantaneous time-dependent frequency shift can be modeled by Equation~(\ref{eq:4}). When the resulting input signal is multiplied by the reference frequency using the lock-in amplifier with its own relative phase $\phi_{LIA}$ and filtered at time constant $\tau$ such that $\frac{1}{\abs{\omega - 2\omega_r}}~>~\frac{\tau}{2\pi}$, the voltage output from the lock-in amplifier is:
\begin{equation*}
\begin{gathered}
V_{LIA}(\Delta_c, t) = \abs{V(\Delta_c, \omega_r)} \times \frac{1}{2}\alpha [E^2_{0}\cos^2(\theta)\cos(\phi_d - \phi_{LIA}) \\
+~E^2_{1}\cos(\phi_d - \phi_{LIA})  \\ 
+~2E_{0}E_{1} \cos(\theta)\cos\big((\omega - 2\omega_r) t - \phi_d - \phi_{LIA}\big)]
\end{gathered}
\end{equation*}

\noindent where $V_{LIA}(\Delta_c, t)$ represents the time-dependent voltage output from the lock-in amplifier. Finally, when terms not dependent on changing signal field $E_{1}$ are removed, we recover that the lock-in amplifier output voltage is proportional to Equation~(\ref{eq:4}) up to an arbitrary phase offset:
\begin{equation}
V_{LIA}(\Delta_c) \propto \frac{1}{2}\alpha E^2_{1} + \alpha E_{0}E_{1}\cos(\theta)
\label{eq:6}
\end{equation}

\noindent meaning that the frequency shift $\Delta_c$ in $\frac{\partial \Delta_c}{\partial V_{rf}}$ can be directly referenced to the Stark shift. The expression for the frequency shift in Equation~(\ref{eq:6}) was then used for the calibration fit done shown in the middle subfigure of Fig.~\ref{fig:6}. Furthermore, as shown at the bottom of Fig.~\ref{fig:6}, this calibration could be used to estimate $E_0\cos{\theta}$ for each signal frequency. The value of $E_0\cos{\theta}$ is treated phenomenologically here, without speculating about its origin. It has been investigated in more depth elsewhere~\cite{schlossberger2025angle,Schlossberger:25}

\begin{figure*}[ht]
    \centering
    \includegraphics[width = 0.6\linewidth]{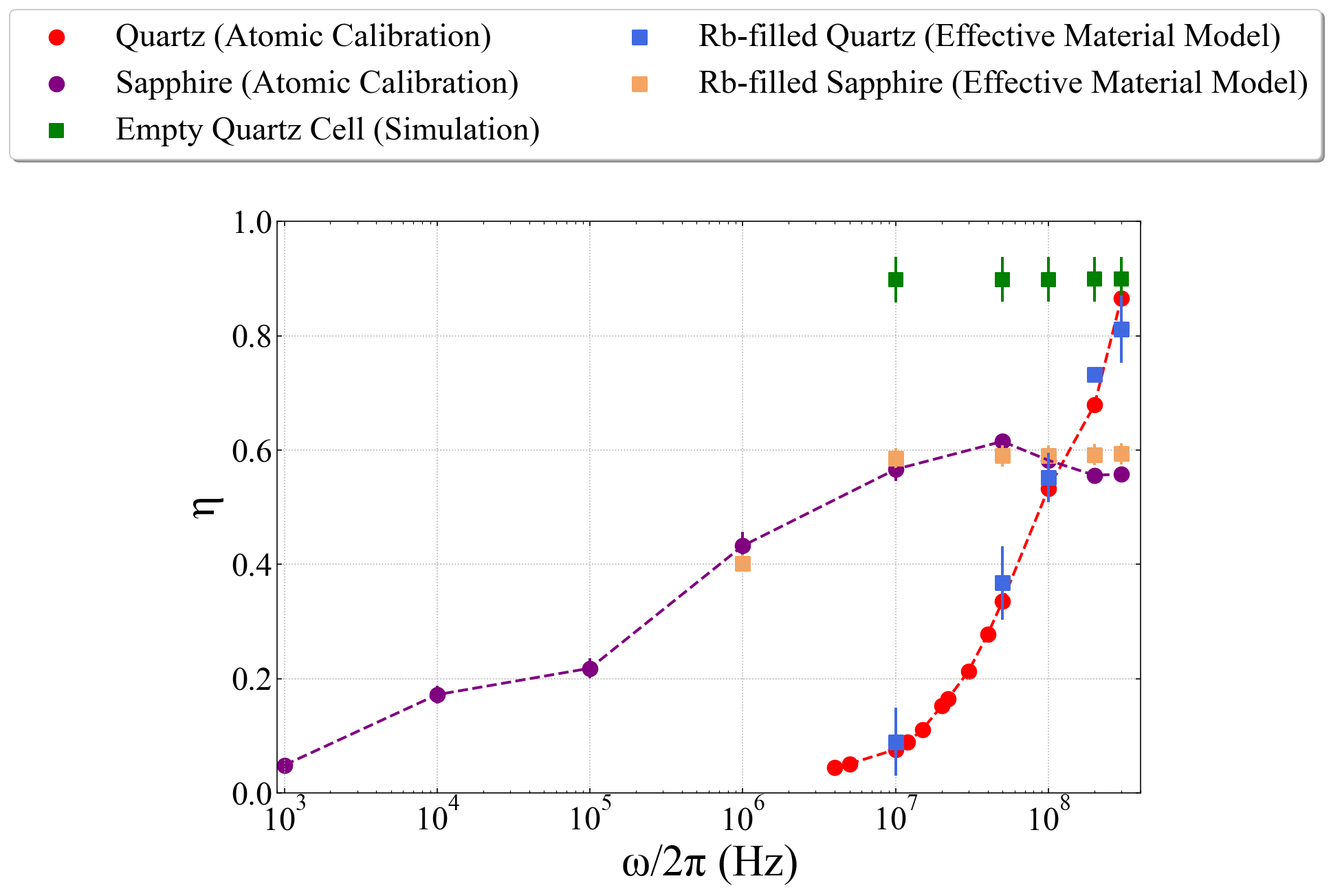}
    \caption{Measures of the frequency-dependent shielding factor $\eta(\omega)$ for atomic calibrations of sapphire (purple) and quartz (red) vapor cells. Also overlaid are simulated conductive screening ratios modeled with extracted material parameters for the sapphire vapor cell (tan), empty quartz vapor cell with base conductivity (green), and quartz vapor cell with effective conductivity (light blue). $\eta = 1$ represents the screening factor of a field interior to an waveguide without a vapor cell. Lines connecting Rydberg-measured points (dashed) are drawn to guide the eye.}
    \label{fig:7}
\end{figure*}

\subsubsection{Screening Measurements}
We now characterize the frequency-dependent shielding ratio $\eta_1(\omega)$ between the free-space electric field $E_{vac}$ and the resulting electric field interior to the vapor cell $E_1$, as in Equation~(\ref{eq:1}). Across both frequency regimes, the calibration fit returns an expression for the slope ($\xi$) of signal field interior to the vapor cell ($E_{1}$) relative to the measured $V_{rf}$:
\begin{equation*}
E_{1}(\omega) = \xi(\omega) V_{rf} 
\end{equation*}

Using this slope $\xi(\omega)$ for each signal frequency, we construct the plot in Fig.~\ref{fig:7}, which displays the expected field screening coefficient from the atomic calibrations for the quartz and sapphire cells. This was done by first evaluating the modeled TEM waveguide for a variety of signal frequencies without a vapor cell, and then normalizing the experimentally measured field from the atomic calibration (represented by $E_{1}$,~$\xi_{1}$) at each frequency against the corresponding field with the empty waveguide (represented by $E_{vac},~\xi_{vac}$):
\begin{equation*}
\frac{E_{1}}{E_{vac}} = \frac{\xi_{1}}{\xi_{vac}} = \eta_1(\omega)
\end{equation*}

\noindent $\xi_{vac}$ was found to range from $\xi_{vac}(2\pi\times$1~kHz) = 20.19~m$^{-1}$~to~$\xi_{vac}(2\pi\times$ 300~MHz) = 25.77~
m$^{-1}$, which can be compared to the infinite parallel plate result of 1/d = 20~$\mathrm{m}^{-1}$ for our septum-ground separation distance of d = 5~cm.

The same was then done for the electric fields simulated in the waveguide with the effective material extraction model to yield simulated shielding ratio $\eta_{sim}(\omega)$. A table of Debye properties found for differing cell conditions, including those without corresponding Rydberg measurements, is given in~\cite{danielTEMpaper}, and partially replicated below. 

\begin{table}[ht]
\begin{tabular}{|c|c|c|c|c|}
\hline
\textbf{Material}       & \textbf{$\epsilon_s$} & \textbf{$\epsilon_{\infty}$} & \textbf{$\frac{\omega_0}{2\pi}$} & \textbf{$\sigma$} \\ \hline
Unfilled Quartz         & --                    & 3.8                       & --                               & --            \\ \hline
$^{87}$Rb-Filled Quartz & 9.6                   & 3.8                       & 113 MHz                          & 0.10 S/m          \\ \hline
Rb-Filled Sapphire      & --                    & 9                         & --                               & 0.001 S/m                 \\ \hline
\end{tabular}
\caption{Debye model parameters from \cite{danielTEMpaper} for simulations of empty quartz cell, a quartz vapor cell filled with Rb$^{87}$, and a Rb-filled sapphire vapor cell. Unfilled entries correspond to unused fitting parameters.}
\label{table0}
\end{table}

Fig.~\ref{fig:7} compares our measured and simulated screening coefficients $\eta_1(\omega)$ and $\eta_{sim}(\omega)$ for the rubidium-loaded quartz vapor cells, which independently verifies the effective material extraction method. While we do not probe the sources of conductive and dielectric contributions to the material properties, we note that they are not characterstic of unfilled cells~\cite{danielTEMpaper}. Prior work has identified these as arising from the dynamical equilibrium of alkali atoms adsorbed on the surface, and more speculatively, collisionally-assisted photo-ionization~\cite{bouchiat1999electrical, vadla1998energy, gallagherbook}. Both of these interactions are dependent on vapor density and cell temperature. The error bars in Fig.~\ref{fig:7} are given by the standard deviation of the $\xi(\omega)$ fit parameter found via the atomic calibrations and the standard deviation of spatially averaged simulated electric field $E_{sim}$ for simulated results. While the quartz cell attenuates strongly for frequencies below 100 MHz, it also allows a larger field fraction to reach the atomic vapor above this crossover frequency compared to the sapphire cell due to its reduced dielectric constant $\epsilon_{\infty}$. We do not report simulation estimates for $\omega<2\pi \times 1$~MHz due to the high noise of the measurement process resulting from degraded performance of the vector network analyzer at lower frequencies. 

\subsection{\label{ssec:sensitivity}Noise-Equivalent Field}

\begin{figure*}[]
\centering
\setkeys{Gin}{width=0.4\textwidth}
\subfloat[
          \label{fig:8a}]{\includegraphics{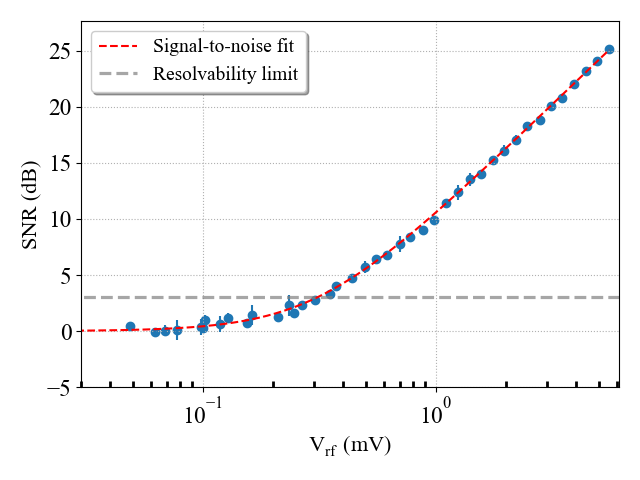}}
\subfloat[
          \label{fig:8b}]{\includegraphics{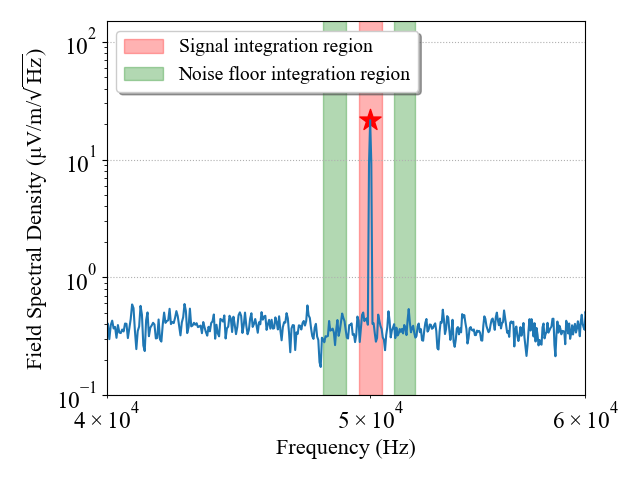}}
          
\subfloat[
          \label{fig:8c}]{\includegraphics{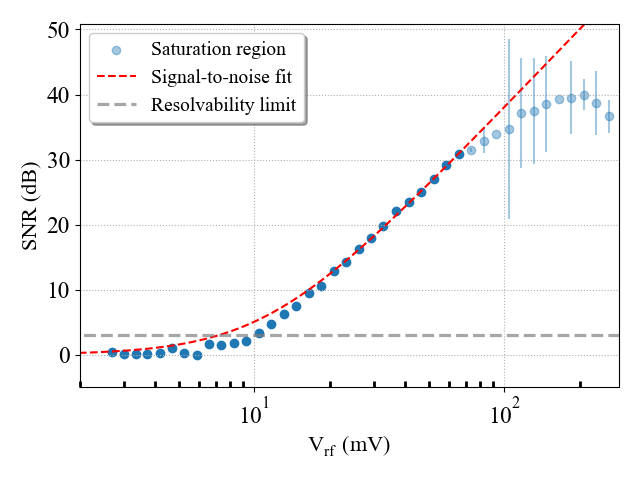}}
\subfloat[
          \label{fig:8d}]{\includegraphics{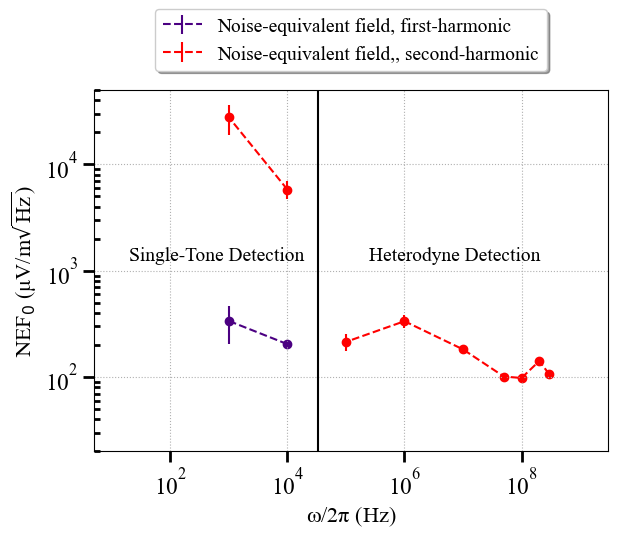}}
\caption{(a) Bandwidth-normalized SNR vs. $V_{rf}$ for $\omega = 2\pi \times 100$~kHz, with overlaid quadratic fit. The intersection of the dashed grey lines is meant to guide the eye to where the noise-equivalent field at the resolvability limit was calculated. (b) Example electric field spectral density $S_E^{(PD)}(\omega)$ for $\omega = 2\pi \times 300$~MHz, displaying the integration region for the applied rf signal in red and the regions averaged for the noise floor in green (c) $2\omega$ bandwidth-normalized SNR vs. $V_{rf}$ for $\omega = 2\pi \times 10$~kHz, with overlaid quadratic fit and saturation region highlighted. The intersection of the dashed grey lines is meant to guide the eye to where the noise-equivalent field at the resolvability limit was calculated. (d) Noise-equivalent field plotted across carrier frequency $\omega$, with noise-equivalent fields measured with heterodyne detection and single-tone detection specified.}
\label{fig:8}
\end{figure*}

An accompanying series of noise-equivalent field ($\mathrm{NEF}_0$) measurements were also taken in the sapphire vapor cell to demonstrate the limits of detection at low frequency, where the noise-equivalent field defining the intrinsic noise floor of the receiver is given by~\cite{santamaria2022comparison}:
\begin{equation*}
\mathrm{NEF}_0 = \frac{|E_1|_{min}}{\sqrt{\Delta f}}
\end{equation*}

\noindent for minimum detectable incident rf signal field amplitude $|E_1|_{min}$ and bandwidth $\Delta f$. The noise-equivalent field and SNR (signal-to-noise ratio) measurements were taken via two separate methods for higher and lower carrier frequencies.

\subsubsection{Time-Averaged Noise-Equivalent Fields ($\mathit{\omega \geq 2\pi\times100}$~kHz)}
The method at higher signal field frequency relies on a signal rf field $E_1$ detuned by 50~kHz from a strong local oscillator rf field $E_{LO}$. The local oscillator field is co-polarized with the rf signal field polarization $\hat{\epsilon}_1$ and the 1257~nm coupler laser is  locked to create a heterodyne "beatnote" when mixed down to baseband by the atomic response~\cite{jingheterodyne}. Modifying the expression for the total AC Stark frequency shift in Equation~(\ref{eq:4}) with an additional local oscillator term of amplitude $E_{LO}$ and phase $\phi_{LO}$ we get~\cite{meyer2021waveguide}:
\begin{equation}
\Delta f_a = -\frac{1}{2}\alpha E_1 E_{LO} \cos(\Delta\omega t + (\phi_1 - \phi_{LO}))
\end{equation}

\noindent where $\Delta\omega = \abs{\omega - \omega_{LO}} = 2\pi\times 50$~kHz is the relative detuning of the two fields (the "beatnote"). The expression is valid for the signal field frequency regime above 100~kHz, $\omega \gg \Delta\omega$. In this regime, the instantaneous bandwidth of the atomic response exceeds the beatnote frequency such that ${2\pi f_{BW}} \gg \Delta\omega$ for instantaneous atomic bandwidth $f_{BW} \approx 150$~kHz. Therefore, all components oscillating faster than the beatnote frequency are time-averaged in the atomic response from the probe transmission~\cite{Yang2024-ph,Bohaichuk2022-hr,jingheterodyne}.

For each noise-equivalent field measurement, the TEM output port voltage $V_{rf}$ was measured with a spectrum analyzer. Then, the average value of the beatnote $\Delta\omega$ in the band-limited power spectrum of the probe transmission photodiode signal ${s}_V^{(PD)}(\Delta\omega)$ with units $V^2$ was evaluated at the heterodyne frequency to measure the total photodiode power spectrum voltage $V_{PD}$ in the beatnote signal:
\begin{equation*}
V_{PD} = \sqrt{\frac{1}{\epsilon}\int^{\Delta\omega+\epsilon/2}_{\Delta\omega-\epsilon/2}{s}_V^{(PD)}(\omega')d\omega'}
\end{equation*}

\noindent for integration bandwidth $\epsilon = 2\pi \times 1$~kHz. As a result, each noise-equivalent field measurement yielded a photodiode voltage to signal voltage calibration $\frac{V_{rf}}{V_{PD}}$. Using the incident electric field $E_{vac}$ inside the TEM waveguide at input signal voltage $V_{rf}$ to further calibrate incident electric field to photodiode power spectrum voltage $V_{PD}$,
\begin{equation*}
\frac{V_{rf}}{V_{PD}} \frac{E_{vac}}{V_{rf}} = \frac{E_{vac}}{V_{PD}}
\end{equation*}

\noindent we extracted an averaged value for the noise floor in the power spectral density in units $[\frac{V}{m~s^{-1/2}}]$~\cite{brown2023very,jingheterodyne}:
\begin{widetext}
\begin{equation}
\label{eq:8}
\mathrm{NEF_0}(\Delta\omega) = \frac{1}{2}\left(\sqrt{\frac{1}{\epsilon}\int^{\Delta\omega+2\epsilon}_{\Delta\omega+\epsilon}{S}_E^{(PD)}(\omega')d\omega'}  + \sqrt{\frac{1}{\epsilon}\int^{\Delta\omega-\epsilon}_{\Delta\omega-2\epsilon}{S}_E^{(PD)}(\omega')d\omega'}~\right)
\end{equation}
\end{widetext}

\noindent where $S_V^{(PD)}(\omega)$ is the photodiode electric field spectral density with units $[\frac{V^2}{s^{-1}}]$, $S_E^{(PD)}(\omega) = (\frac{E_{vac}}{V_{PD}})^2S_V^{(PD)}(\omega)$ is the photodiode electric field spectral density with units $[\frac{V^2}{m^2s^{-1}}]$, and $\mathrm{NEF_0}(\Delta\omega)$ is defined as the heterodyne noise-equivalent field. From here, noise-equivalent fields are plotted and expressed in units $[\frac{V}{m~Hz^{1/2}}]$. An example of the heterodyne power spectral density and the integrals above are demonstrated in Figure~\ref{fig:8b}. The integration regions for the noise floor were chosen arbitrarily to sample regions both above and below the heterodyne beatnote to compensate for any frequency dependence in the noise. The corresponding dimensionless SNR was also calculated by taking the ratio of the power spectral density in electric field at the heterodyne frequency over $\mathrm{NEF_0}$:
\begin{equation}
\label{eq:9}
\epsilon \times \mathrm{SNR_{\Delta\omega}}~=~\frac{\int^{\Delta\omega+\epsilon/2}_{\Delta\omega-\epsilon/2}{S}_E^{(PD)}(\omega')d\omega'}{\mathrm{NEF_0}({\Delta\omega)^2}}
\end{equation}

\noindent where the SNR is evaluated as the signal-to-noise ratio in a 1~Hz bandwidth. In short, the voltage whose amplitude corresponds to the known applied field was measured at the beatnote on the photodiode power spectrum and defined our noise floor in field units, allowing us to derive a noise-equivalent field value.

\subsubsection{Adiabatic Noise-Equivalent Fields  ($\mathit{\omega < 2\pi\times100}$~kHz)}
In the previous section, for $\omega > 2\pi\times 100$~kHz, $\omega \gg \Delta\omega$ and $2\pi f_{BW}~\gg~\Delta\omega$ for heterodyne beatnote frequency $\Delta\omega = 2\pi\times 50$~kHz. In the regime below 100~kHz where the first assumption no longer holds, no local oscillator field was applied. Furthermore, as a consequence of the second assumption, the frequency components of the probe transmission oscillating at the signal frequency and its second harmonic no longer oscillate faster than the 4-level system's dynamic response frequency. In this "adiabatic" regime, the frequency shift of the dressed states follows the instantaneous value of the incoming fields, similarly to the "adiabatic" calibration regime mentioned prior. Expanding Equation~(\ref{eq:4}) without any assumed time-averaging, we find:
\begin{align*}
    \begin{gathered}
\Delta f_a = -\frac{1}{2}\alpha\Biggl(E_0^2 \cos^2(\theta) + \frac{1}{2} E_1^2 + \frac{1}{2} E_1^2\cos(2\omega t + 2\phi_1) \\ +~2 E_0 E_1 \cos(\omega t + \phi_1) \cos(\theta)\Biggr)
    \end{gathered}
\end{align*}

It follows that the power spectrum of the probe transmission  includes two time-dependent components: one oscillating at fundamental $\omega$ and one at second harmonic $2\omega$. This results in two measures of noise-equivalent field: a "first-harmonic" $\mathrm{NEF_0}$ using the power for the $\omega$ peak that is proportional to the product of the dc bias field amplitude and the rf signal field amplitude ($E_0 E_1)$, and a non-linear $\mathrm{NEF_0}$ using the $2\omega$ peak that is proportional to the square of the rf signal field amplitude ($E_1^2)$. Then, similarly to the analysis process for the heterodyne case, we can evaluate both $\mathrm{NEF_0}(\omega)$ and $\mathrm{NEF_0}(2\omega)$ in the frequency domain. 

For these measurements, the TEM output port voltage $V_{rf}$ was measured with an oscilloscope since the signal frequencies fell below the spectrum analyzer operating range. Then, for the second-harmonic noise-equivalent field, the $2\omega$ peak in the photodiode power spectral density was integrated to find a photodiode voltage to rms signal voltage calibration factor $\frac{V_{2\omega}}{V_{rf}}$:
\begin{equation*}
\frac{V_{2\omega}}{V_{rf}} =  \frac{1}{V_{rf}}\sqrt{\int_{2\omega-\epsilon/2}^{2\omega+\epsilon/2}S_V^{(PD)}(\omega')~d\omega'}
\end{equation*}

\noindent for integration bandwidth $\epsilon = 2\pi \times 1$~kHz and photodiode voltage spectral density $S_V^{(PD)}(\omega')$ with units $[\frac{V^2}{s^{-1}}]$. The effective material extraction model at each frequency allows us to convert the photodiode power spectral density to a "second-harmonic" electric field $\mathrm{NEF_0}(2\omega)$ in terms of incident electric field $E_{vac}$, analogous to Equation~(\ref{eq:8}).

This can similarly be done for the first harmonic $\omega$ in the probe response to return a "first-harmonic" $\mathrm{NEF_0}(\omega) \propto E_0E_1$, again analogously to Equation~(\ref{eq:8}). The SNR could then be evaluated for both second- and first-harmonic measurements to yield $\mathrm{SNR_{2\omega}}$ and $\mathrm{SNR_{\omega}}$ analogously to Equation~(\ref{eq:9}).

\begin{table}[]
\begin{tabular}{|c|c|}
\hline
\textbf{Frequency (MHz)} & \textbf{$\mathrm{NEF}_0$($\mu V/m\sqrt{Hz}$)} \\ \hline
\textbf{300}                      & 106 (4)                                             \\ \hline
\textbf{200}                      & 141 (4)                                             \\ \hline
\textbf{100}                      & 98 (8)                                             \\ \hline
\textbf{50}                       & 101 (5)                                             \\ \hline
\textbf{10}                       & 182 (8)                                             \\ \hline
\textbf{1}                        & 335 (11)                                            \\ \hline
\textbf{0.1}                      & 214 (10)                                            \\ \hline
\end{tabular}
\caption{Heterodyne noise-equivalent fields and corresponding standard deviations versus signal frequency, with uncertainties in parentheses}
\label{table1}
\end{table}

\begin{table}[]
\begin{tabular}{c|cl|}
\cline{2-3}
\multicolumn{1}{l|}{}                & \multicolumn{2}{c|}{\textbf{$\mathrm{NEF}_0$ ($mV/m \sqrt{Hz}$)}} \\ \hline
\multicolumn{1}{|c|}{\textbf{Frequency (MHz)}} & \multicolumn{1}{c|}{\textbf{2$\bm{\omega}$}} & \multicolumn{1}{c|}{\textbf{$\bm{\omega}$}} \\ \hline
\multicolumn{1}{|c|}{\textbf{0.01}}  & \multicolumn{1}{c|}{5.8  (1.1)}    &  \multicolumn{1}{c|}{0.2 (0.009)}     \\ \hline
\multicolumn{1}{|c|}{\textbf{0.001}} & \multicolumn{1}{c|}{27.7 (8.7)}   & \multicolumn{1}{c|}{0.34 (0.13)}   \\ \hline
\end{tabular}
\caption{Second- and first-harmonic noise-equivalent fields and corresponding standard deviations versus signal frequency, with uncertainties in parentheses}
\label{table2}
\end{table}

\begin{figure}[]
    \includegraphics[width = \linewidth]{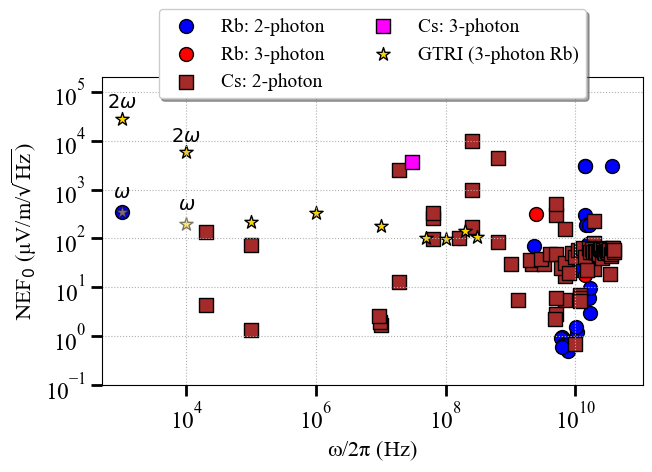}
    \caption{Noise-equivalent fields from across the scientific literature for experimental designs not using a field-enhancing structure in the vicinity of the cell~\cite{sedlacek2013polarization,Sedlacek2012-jc,Borowka2024-ku,jingheterodyne,kumar2017atom,gordon2019weak, Prajapati2021-gi,Hu2022-qe,Holloway2022-zo,Cai2023-wg,Downes2020-hk,Yang2024-sl,Mao2024-lk,Sandidge2024-pb,photonshotnoise-shaffer2017,Liang2025-ay,Cai2022-fm,Romalis2024-hh,nowosielski2023superheterodyne,She2024-vj,Liu2022-yc,Bohaichuk2022-hr,Dixon2023-qh,Berweger2023-fx,Yang2024-ph,Xie2025-gz,Hu2025-bw}, where "Rb" and "Cs" indicate noise-equivalent field results for 2- and 3-photon schemes in rubidium and cesium vapor cells. Star markers labeled "GTRI (3-photon Rb)" indicate $\mathrm{NEF}_0$ published in this paper, and markers for noise-equivalent fields taken below 100~kHz include annotations designating whether they are second-harmonic ($2\omega$) or first-harmonic ($\omega$) $\mathrm{NEF}_0$}
    \label{fig:9}
\end{figure}

\subsubsection{Characteristic Noise-Equivalent Fields}
For each signal frequency there are two distinct regions in terms of the scaling of atomic response with field amplitude: a "quadratic" region wherein the atomic response scales quadratically with applied field, and a "saturation" region where the atomic response is saturated and exhibits reduced scaling with signal field. To determine characteristic $\mathrm{NEF_0}$ at the resolvable field limit for each signal frequency $\omega$, a fit was done over the measured SNRs according to a quadratic function (as seen in Fig.~$\ref{fig:9}$):
\begin{gather*} 
\mathrm{SNR}(V_{rf}) = 1 + aV_{rf}^2 \\  
0 \leq V_{rf} \leq V_{sat}
\end{gather*}

\noindent where $V_{sat}$ is the rf rms voltage applied at the onset of the saturation region, and $a$ is the fit parameter in the quadratic region. Then, the characteristic noise-equivalent field was taken to be the value of the quadratic fit function at $V_{res}$, where $V_{res}$ is the rf rms voltage applied at the field resolvability limit SNR~=~3~dB. Table~\ref{table1} displays the measured heterodyne noise-equivalent fields for $\omega \geq 2\pi\times100$~kHz. Table~\ref{table2} displays the measured noise-equivalent fields versus signal frequency for second- and first-harmonic noise-equivalent field measurements for $\omega < 2\pi\times100$~kHz, respectively.

\section{\label{sec:out}Discussion}
Fig.~\ref{fig:8d} shows the resulting noise-equivalent fields versus signal frequency. Each of the GTRI $\mathrm{NEF_0}$ are referenced to the free-space field amplitude, and not to the amplitude measured in the vapor cell. We achieve noise-equivalent fields of 106 (4)~$\mathrm{\frac{uV/m}{\sqrt{Hz}}}$ at $\omega = 2\pi \times 300$~MHz and 27.7~(8.7)~$\mathrm{\frac{mV/m}{\sqrt{Hz}}}$ second-harmonic noise-equivalent field at $\omega = 2\pi \times 1$~kHz, which can be further improved by using the first-harmonic noise-equivalent field of 0.34~(0.13)~$\mathrm{\frac{mV/m}{\sqrt{Hz}}}$. 

Fig.~\ref{fig:9} displays a comparison of the noise-equivalent fields measured in this paper against a variety of noise-equivalent fields measured in the literature~\cite{sedlacek2013polarization,Sedlacek2012-jc,Borowka2024-ku,jingheterodyne,kumar2017atom,gordon2019weak,Prajapati2021-gi,Hu2022-qe,Holloway2022-zo,Cai2023-wg,Downes2020-hk,Yang2024-sl,Mao2024-lk,Sandidge2024-pb,photonshotnoise-shaffer2017,Liang2025-ay,Cai2022-fm,Romalis2024-hh,nowosielski2023superheterodyne,She2024-vj,Liu2022-yc,Bohaichuk2022-hr,Dixon2023-qh,Berweger2023-fx,Yang2024-ph,Xie2025-gz,Hu2025-bw}. The figure displays favorable second-harmonic and first-harmonic noise-equivalent fields for the sapphire vapor cell at low rf frequency without any field enhancement or resonant structure.

In this work, we have measured the low-frequency screening factors of fused quartz and sapphire vapor cells in the frequency range of 1~kHz to 300~MHz and shown them to be consistent with an all-electrical measurement and HFSS simulation produced elsewhere~\cite{danielTEMpaper}. Future work includes utilizing the outputs of the effective material extraction model and the DC bias field shift imposed by charges on the walls to image the local density and distribution of adsorbates in vapor cells manufactured from silica glasses with substantial surface interactions with alkali metals.

\section{\label{sec:ack}Acknowledgements}
The authors thank C. D. Herold and K. R. Brown for comments and careful reading of the manuscript. 

\bibliography{apssamp}

\end{document}